\documentclass{aa}
\usepackage{graphicx}
\usepackage{color}
\usepackage{txfonts}
\usepackage[thinspace,mediumqspace,squaren]{SIunits}
\begin{document}
\title{Inferring the interplanetary dust properties}

   \subtitle{From remote observations and simulations}

   \author{J. Lasue\inst{1}          
          \and A.C. Levasseur-Regourd\inst{1,2}
          \and N. Fray \inst{3}
          \and H. Cottin \inst{3}
   }

   \offprints{J. Lasue}

   \institute{Service d'a\'eronomie-IPSL-CNRS, UMR 7620, Route des G\^atines BP-3,  
Verri\`eres-le-Buisson, F-91371 France\\
     \email{jeremie.lasue@aerov.jussieu.fr}
     \and
     Universit\'e Pierre et Marie Curie-Paris6, Service d'a\'eronomie UMR 7620, Paris, F-75005 France ; \\
     \email{chantal.levasseur-regourd@aerov.jussieu.fr}
     \and 
     Laboratoire Interuniversitaire des Syst\`emes Atmosph\'eriques, UMR 7583, 
     Universit\'es Paris 7 et Paris 12, 61 av. du G\'en\'eral de Gaulle, Cr\'eteil, F-94010 France\\
     \email{fray@lisa.univ-paris12.fr;\, cottin@lisa.univ-paris12.fr}
     %             \thanks{The university of heaven temporarily does not
     %                     accept e-mails}
             }

   \date{}

% \abstract{}{}{}{}{} 
% 5 {} token are mandatory
 
  \abstract
   {Since in situ studies and interplanetary dust collections only provide a spatially limited
amount of information about the interplanetary dust properties, it is of major importance to complete
these studies with properties inferred from remote observations of light scattered and emitted, with 
interpretation through simulations. 
}
   {Physical properties of the interplanetary dust in the near-ecliptic symmetry surface, 
such as the local polarization, temperature and composition, 
together with their heliocentric variations, may be derived from scattered and emitted light 
observations, giving clues to the respective contribution of the particles sources.}
   {A model of light scattering by a cloud of solid particles constituted by spheroidal grains 
and aggregates thereof is used to interpret the local light scattering data.
Equilibrium temperature of the same particles allows us 
to interpret the temperature heliocentric variations.}
   {A good fit of the local polarization phase curve, $P_{\alpha}$, near 1.5~AU from the Sun
is obtained for a mixture of  silicates and more absorbing  organics material ($\approx$40 \% in mass)
and for a realistic size distribution typical of the interplanetary dust 
in the \unit{0.2}{\micro\metre} to \unit{200}{\micro\metre} size range.
The contribution of dust particles of cometary origin is at least 20\% in mass. 
The same size distribution of particles gives a solar distance, $R$, dependence of the temperature 
in $R^{-0.45}$ different than the typical black body behavior.
The heliocentric dependence of $P_{\alpha=\unit{90}{\degree}}$ is interpreted as a
progressive disappearance of solid organics (such as HCN polymers or amorphous carbon) towards the Sun.
}



   \maketitle
%

%________________________________________________________________
\section{Introduction}

The description of the particles constituting the interplanetary dust cloud (IDC) 
in terms of morphology, porosity, 
size distribution and complex refractive indices is a clue to their origin and evolution.
Information can be retrieved through (a few) in situ studies (see e.g. Jessberger et al. \cite{ekj_ts01})
and through remote observations of the light scattering properties 
(brightness and polarization of solar light scattered by the dust particles in the visible domain) 
and emissivity (in the infrared spectrum) of the dust cloud, see e.g. Levasseur-Regourd et al. (\cite{aclr_mc99}). 
Photometric measurements integrate along the line of sight all the local contributions 
emitted and scattered by the dust. 
Consequently techniques of inversion, such as the nodes of lesser uncertainty method 
(see e.g. Dumont \& Levasseur-Regourd \cite{rd_aclr88}, Levasseur-Regourd et al. \cite{aclr_im01}
and references therein) are needed to retrieve the local properties. 
All the values given in the following text correspond 
to bulk values deduced from inversion methods over elementary volumes.

Light scattering numerical simulations can be used to derive physical properties 
of clouds of dust particles, typically of comet origin 
(see e.g. Levasseur-Regourd et al. \cite{aclr_tm07} and references therein).
Realistic light scattering models for a distribution of particles constituted of a mixture 
of spheroidal grains and aggregates of small spheroids have already been used to 
derive information about the composition and size distribution 
(lower and upper cut-off, power law coefficient, proportion of absorbing and non-absorbing 
material and proportion of aggregates) in the case of comet Hale-Bopp dust polarimetric observations 
(Lasue \& Levasseur-Regourd \cite{jl_aclr06}).

This study presents the results obtained by applying an irregular particles cloud model 
in the case of the interplanetary dust cloud observations to estimate the physical properties 
of the size distribution and the proportion of fluffy particles. It tentatively indicates 
the relative contribution of particles from cometary and asteroidal origins. 
In the next two paragraphs, clues to the properties of the interplanetary dust cloud and 
source particles are reviewed.
In the last two paragraphs, the emitted light and local temperature properties of the cloud are analyzed
through our model.

%________________________________________________________________
\section{Local properties of the interplanetary dust cloud}

\subsection{Scattered light}

The local albedo of the interplanetary dust cloud, $A$,  
follows approximately $A = ( 0.07 \pm 0.03 )  \, R^{-0.34 \pm 0.05}$ 
in the near-ecliptic symmetry surface as a function of the solar distance, $R$, between 1.5 and 0.5~AU 
as deduced from the brightness observations in the visible 
(Dumont \& Levasseur-Regourd \cite{rd_aclr88},
Levasseur-Regourd et al. \cite{aclr_im01}).

The local polarization, $P$, cannot be estimated at all the phase angles and 
solar distances. With the nodes of lesser uncertainty method, two zones can be described. 
The radial node gives information on the polarization at \unit{90}{\degree} of phase angle with 
a varying solar distance, whereas the martian node located at 1.5~AU from the Sun gives 
the variation of $P$ with the phase angle (Dumont \& Levasseur-Regourd \cite{rd_aclr88}).
The deduced phase curve presents shape similar to the ones observed for comets  and other dusty objects of the 
Solar System, with large error bars due to inversion methods at a single wavelength 
\unit{\lambda=550}{\nano\metre} and
with a shape typical of the interaction of light with irregular particles of a size 
comparable to the wavelength (see e.g.  Mann \cite{im92}, Lumme \cite{kl00},
Levasseur-Regourd \& Hadamcik \cite{aclr_eh03}).
The phase curve is smooth with a small negative branch below 
the inversion angle, $\alpha_0 \approx \unit{15}{\degree}\pm \unit{2}{\degree}$
and a large positive branch, with a value $P_{\unit{90}{\degree}}\approx 30 \% \pm 3\%$.

From the observational data, no significant variation of  $P$ with the wavelength, $\lambda$, 
can be noticed, as presented by Leinert et al. (\cite{cl_sb98}).
However a heliocentric dependence of $P_{\unit{90}{\degree}}$ between 1.5 and 0.5~AU
can be pointed out (Lumme \cite{kl00}, Levasseur-Regourd et al. \cite{aclr_im01}):
\begin{equation}
P_{\unit{90}{\degree}} = ( 0.30 \pm 0.03 ) \, R^{0.5 \pm 0.1}
\label{var_P}
\end{equation}
This relation suggests that a change either in the physical properties or in the chemical composition 
of the particles is required to interpret the observed solar distance dependence of the IDC local polarization.

\subsection{Temperature}

Although the absolute temperature values may not be accurately determined, 
the variation of local temperature of the IDC with the solar distance is well constrained. 
The dependence of $T$ with $R$  follows closely a power law, the  exponent, $t$, 
(hereafter called temperature-distance factor),
of which is less steep than the one expected from a black body ($T$ proportional to $\frac{1}{\sqrt{R}}$):
\begin{equation}
T = ( \unit{250 \pm 10}{\kelvin} )  \, R^{-0.36 \pm 0.03}
\label{var_T}
\end{equation}
This relation is approximately valid between 1.5 and 0.5~AU as deduced from the 
infrared observations (Dumont \& Levasseur-Regourd \cite{rd_aclr88}, Reach \cite{wtr91},
Renard et al. \cite{jbr_aclr95}, Levasseur-Regourd et al. \cite{aclr_im01}).

\section{Clues to the properties of source particles}

The IDC, as probed around 1~AU, originates  from at least three different sources: 
fluffy and easily fragmenting particles  of cometary origin (see e.g. Whipple \cite{flw51});
 asteroidal dust made of compact particles produced by asteroidal shattering 
(see e.g. Jessberger et al. \cite{ekj_ts01}); 
and to a lesser extent elongated submicron interstellar grains passing through the solar system and a priori
not directly linked to the primordial grains that formed the solar system 
and grains coming from the Jovian system (see e.g. Gr\"un et al. \cite{eg_haz93}).

\subsection{Sizes of the particles}

The size distribution of compact or fluffy particles resulting from particle-particle collisions 
is expected to follow a power law,
$a^{s}$, where the effective diameter, $a$, is the diameter of the sphere with a volume equivalent to
the one of the irregular particle, and where $s$ is equal to $-3$ from theoretical
calculations (Hellyer \cite{bh70}) and is between $-3.5$ and $-3$ from experimental
simulations (see e.g. Mukai et al. \cite{tm_jb01}). 
Observations of the solid component of the cometary com\ae\, have shown that such a coefficient
is close to $-3$ (e.g. $-3.1 \pm 0.3$ from Rosetta observations of 
comet 9P/Tempel 1 after Deep Impact event by Jorda et al. \cite{lj_pl07}).
In situ measurements within the IDC suggest $s$ to be around $-3$ for particles smaller than \unit{20}{\micro\metre}
and equal to -4.4 for larger particles (Gr\"un et al. \cite{eg_mb01}).

Interplanetary Dust Particles (IDPs) collected in the Earth stratosphere typical size
ranges from \unit{5}{\micro\metre} to \unit{25}{\micro\metre} (Jessberger et al. \cite{ekj_ts01}). 
For the IDPs presenting an aggregated structure and possibly originating from comets, 
the constituent grains have a mean size around \unit{0.3}{\micro\metre}. 
Evidence for such a fluffy structure has also been shown from the study of 
the aluminium craters and aerogel penetration tracks from the Stardust samples (H\"orz et al. \cite{fh_rb06}).
The larger micrometeorites, collected in the polar regions or at the bottom of the oceans,  
have sizes ranging from \unit{20}{\micro\metre} to \unit{1}{\milli\metre}, and because of their large size 
are usually compact due to melting during their entry through the Earth's atmosphere
(Engrand \& Maurette \cite{ce_mm98}).

\subsection{Optical constants of the particles}

Studies of comets, IDPs and micrometeorites have shown the predominance of 
silicates and ``CHON'' materials in the composition of extraterrestrial particles (Hanner \& Bradley \cite{msh_jpb05}).
Precise spectroscopic studies of silicates have shown the presence of both amorphous and 
crystalline silicates consisting of olivine or pyroxene that could be explained by 
radial mixing of the elements in the early solar nebula (Bockel\'ee-Morvan et al. \cite{dbm_dg02}). 
Models of comets (Wooden et al. \cite{dhw_deh99}, Hayward et al. \cite{tlh_msh00}) as well as 
studies of fluffy IDPs (see e.g. Bradley et al. \cite{jpb_hjh92}) have shown 
that the main constituent of the silicates was actually Mg-rich pyroxene with a contribution between 60\% and 90\% 
and an absorption between 0.001 and 0.01 in the visible domain near \unit{550}{\nano\metre} 
(as expressed in terms of the imaginary part, $k$, of the optical index $m=n+$i$k$, Dorschner et al. \cite{jd_bb95}). 
The recent mineralogical studies of the Stardust  samples have also confirmed a variety of olivine and pyroxene
silicates in various crystallization states as the main component (Zolensky et al. \cite{mez_tjz06}).
In the visible domain, the crystalline silicates present approximately the same optical indices  
than the amorphous, glassy silicates with an absorption $k \approx 0.0001$
 (J\"ager et al. \cite{cj_fjm98}, Lucey \cite{pgl98}).
It may thus be assumed that amorphous and crystalline silicates have similar optical indices in the visible.
In this study, the complex index of silicates will be assumed to be the one of  Mg-rich pyroxene
(typically  Mg$_x$Fe$_{1-x}$SiO$_3$ with $0.8<x<0.7$ 
as measured in laboratory by Dorschner et al. \cite{jd_bb95}), 
as it represents most of the observed silicates (see section 4.1 for more details).

The ``CHON'', or organics component of the particles is a more absorbing material composed of lighter elements 
that could appear in cometary dust particles from the heavy radiation processing of ices and 
light elements deposited on the grains. 
Attemps to characterize the nature of the carbonaceous fraction in IDPs suggest that amorphous carbon 
is dominant (Keller et al. \cite{lpk_klt94}). However, recent studies suggest that one-half of the carbon contained 
in typical IDPs is present in the form of organic carbon (Flynn et al. \cite{gjf_lpk04}).
Laboratory experiments of radiation processing of ices have been shown to produce 
absorbing organics material with a high real part of the optical index (close to 2) and imaginary part 
between 0.1 to 0.4, see e.g. Jenniskens (\cite{pj93}) and Li \& Greenberg (\cite{li_jmg97}). 
Studies of deposited graphite gave optical index values in the visible of 
the order of $1.71+ \rm{i}\,0.1$ (Papoular et al. \cite{rp_jb93}).
The amorphous carbon presents a very high  absorption with an optical index 
in the visible of $1.88+ \rm{i}\,0.71$ (Edoh \cite{oe_83}).
In this study, the complex index of organics material taken from 
Li \& Greenberg (\cite{li_jmg97}) is assumed to reproduce the behaviour 
of irradiated cometary ices, and the complex index of amorphous carbon 
is taken from Edoh (\cite{oe_83}).

The spectrum of the zodiacal light is similar to the solar spectrum
(see e.g. Levasseur-Regourd et al. \cite{aclr_im01}) 
and peaks around \unit{550}{\nano\metre}. As a first approximation, 
for the calculations of light scattering, the complex index of the particles will be taken 
around this particular wavelength. However since 
equilibrium temperature calculations take into account the infrared emission of the particles,
the variation of the complex index with the wavelength will be taken into account for these calculations.

\subsection{Shapes of the particles}

The IDC particles ejected by comets are probably very irregular with both compact and aggregated ones
as shown from light scattering studies (Lasue \& Levasseur-Regourd \cite{jl_aclr06})
and the study of foil craters and aerogel penetration tracks of Stardust (H\"orz et al. \cite{fh_rb06}).
Following the ``bird-nest'' model proposed by Greenberg \& Hage (\cite{jmg_jih90}), 
the aggregates are assumed to be built of spheroidal grains with an axis ratio around 2. 
Such grains might be expected to have a layered structure such as a silicates-core covered by an organics-mantle, 
with a material ratio similar to what is expected from protosolar clouds models. 
These grains of interstellar origin may be used to model the IDC interstellar component.

The IDC particles resulting from asteroidal collisions are probably compact.
They may be in a first approximation represented 
by prolate spheroids with an axis ratio less than 2 as estimated from IDPs
collection analysis (Jessberger et al. \cite{ekj_ts01}) and from collisions experimental
simulations (Mukai et al. \cite{tm_jb01}). 
 More specifically, the study of Na desorption on meteoroids by 
Kasuga et al. (\cite{tk_ty06}) has shown that the meteoroids near 1~AU can be large, compact particles. 

Finally, following the ``bird-nest'' model, 
the interstellar grains present in the interplanetary medium are 
assumed to be similar to the constituent grains of the aggregated particles 
described above.

%__________________________________________________________________
%\subsection{Method}

\section{Simulation of the scattered light}

The variations of the local properties of the dust cloud with the solar distance
 point out to spatial and temporal evolutions of the particles, the small particles being ejected by radiation 
pressure and the largest ones spiralling slowly towards the Sun under Poynting-Robertson effect
(see e.g. Dermott et al. \cite{sfd_kg01}).
We shall now  interpret the light scattered and emission results in terms of properties of the particles 
through an IDC model using a mixture of fluffy and compact particles composed of 
 silicates and (more absorbing) organics material and possibly amorphous carbon
with a model of the IDC including fluffy as well as compact particles constituted 
from silicates and more absorbing organics material.

\subsection{Principle of the calculations}

The light scattering computations are performed for a size distribution of spheroids and aggregates 
thereof following a size distribution similar to the one proposed by Gr\"un et al. (\cite{eg_mb01}) for the 
in situ measurements of the IDC particles, with two different slopes. 
The power law coefficient, $s$, is around $-3$ 
for effective diameters, $a$, smaller than  \unit{20}{\micro\metre}  and steeper ($s=-4.4$) for 
larger effective diameters.

\begin{figure}
\centering
\includegraphics[width=0.8\linewidth]{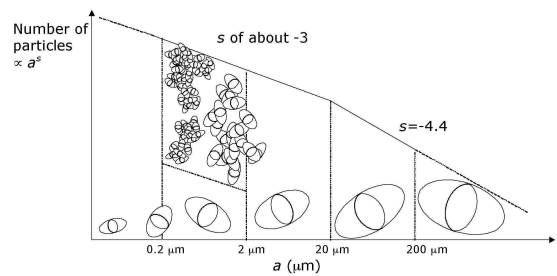}
\caption{Size distribution of prolate spheroids (from effective diameter \unit{a=0.2}{\micro\metre} to \unit{200}{\micro\metre}) 
and aggregates thereof (up to 256 spheroids from \unit{a=0.2}{\micro\metre} to \unit{2}{\micro\metre}).}
\label{sdist}
\end{figure}

Light scattering by prolate spheroids (axis ratio of 2) is calculated by a code adapted from T-matrix for small spheroids 
(\unit{a<3.5}{\micro\metre}, Mishchenko \& Travis \cite{mim_ldt98}) and by ray-tracing for large spheroids 
(\unit{a>3.5}{\micro\metre}, Macke \& Mishchenko \cite{am_mim96}). 
Contribution of Ballistic Cluster-Cluster Aggregates and Ballistic Particle-Cluster Aggregates 
(BCCA and BPCA as defined in Meakin \cite{pm83})  of up to 256 spheroidal grains calculated with 
the Discrete Dipole Approximation (Draine \& Flatau \cite{btd_pjf00})
is taken into account in the cloud for effective diameters of the aggregates, 
$a$, between \unit{0.2}{\micro\metre} and \unit{2}{\micro\metre} due to computation limitations.
Fig.~\ref{sdist} summarizes and illustrates such a size distribution with the mixture of particles.

The brightness, $Z$, and its two polarized components are calculated by integration of the incident 
light intensity $Z_{\rm{inc}}$ over the size distribution, $\Gamma(a)$,
of the dust particles and their scattering cross section $\sigma_{\rm{sca}}(a,\alpha,\lambda)$ at a given 
phase angle $\alpha$.
\begin{equation}
Z(\alpha,\lambda)=\frac{\int_0^{\infty}Z_{\rm{inc}}(\lambda)\sigma_{\rm{sca}}(a,\alpha,\lambda)\,\Gamma(a)\,da}
{\int_0^{\infty}\sigma_{\rm{sca}}(a,\alpha,\lambda)\,\Gamma(a)\,da}
\end{equation}

Details on the calculations together with results on comet Hale-Bopp 
coma composition are given in Lasue \& Levasseur-Regourd (\cite{jl_aclr06}).
In that paper, the optical properties of particles with a silicates-core surrounded by an organics-mantle
(with expected protosolar abundances) are shown to follow closely the ones of pure organics particles. 
Fitting the observational data in terms of silicates and organics particles mixture 
allows us to estimate the range of the carbonaceous material percentage in mass  of the IDC,
taking into account the fact that organics particles can actually embed silicates material.

Since the light scattered by the particles reproduce the solar spectrum in the visible, 
for the following calculations we have taken only the complex index of the material near 
\unit{550}{\nano\metre} as a first approximation of the complex index of the particles.
The optical indices are taken to be those of Mg-rich pyroxene 
($1.62+ \rm{i}\,0.003$ at \unit{\lambda=550}{\nano\metre}, 
for Mg$_x$Fe$_{1-x}$SiO$_3$ with $0.8<x<0.7$, Dorschner et al. \cite{jd_bb95}), 
and refractive organics material obtained through radiation processing of light elements
($\approx 1.88+\rm{i}\,0.1$ at \unit{\lambda=550}{\nano\metre},  Li \& Greenberg \cite{li_jmg97}).

\subsection{Results for the scattered light}

\begin{figure}
\centering
\includegraphics[width=0.8\linewidth]{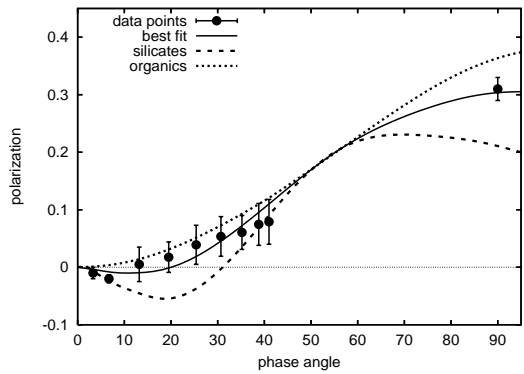}
\caption{Zodiacal dust polarization data ($\bullet$) around 1.5~AU near the symmetry surface
and phase curves calculated at \unit{550}{\nano\metre}. 
Curves are plotted for a realistic interplanetary dust size distribution
of silicates (dashed line), organics (dotted line)  and  a mixture with $\approx 40$ \% of 
organics in mass (thin solid line). Data points are adapted from Levasseur-Regourd et al. \cite{aclr_im01} and 
references therein. The error bars are shown whenever greater than 0.5\%.}
\label{fit_Palpha}
\end{figure}

Fig.~\ref{fit_Palpha} presents the phase curves calculated for the above described IDC size distribution.
The phase curves obtained for silicates and (absorbing) organics particles are significantly different 
from each other and from the data points derived from the observations.
The organics particles phase curve has a high value $P_{\unit{90}{\degree}}$ around 38\% 
and does not show any negative branch,
 in agreement with previous calculations on absorbing aggregates (Kimura \cite{hk01}),
whereas the silicate particles phase  curve presents a low value $P_{\unit{90}{\degree}}$ around 20\% and a value $\alpha_{0}$ 
around $\unit{30}{\degree}$. Contribution of both absorbing and less absorbing materials is thus expected 
to reproduce satisfyingly the IDC composition around 1.5~AU.

As shown in Fig.~\ref{fit_Palpha}, a good fit of the data retrieved 
for the polarization phase curve of the IDC near 1.5~AU 
is  obtained for the above mentioned realistic size distribution 
with a small particles power law coefficient $s=-3.0$ ($s=-4.4$ for 
large particles as previously defined), effective diameter of 
\unit{0.22}{\micro\metre} for the lower cutoff, and of \unit{200}{\micro\metre} for the upper cutoff, and a 
silicates--organics mixture with about 50\% of organics particles in mass.
Taking into account the observational error bars, the mass of these organics particles
is actually comprised between 40\% and 60\%.  
When taking into account the fact that the organics particles in the IDC might embed silicates 
cores, typically with the cosmic abundances considered in Greenberg \& Hage (\cite{jmg_jih90}) model, 
meaning that these particles actually contain less organic material than is visible, 
the total amount of organic material can be estimated to be comprised between 20\% and 60\% in mass. 
The amount of aggregates (probably unfragmented  particles of cometary origin) is of about 20\% in mass.

\begin{figure}
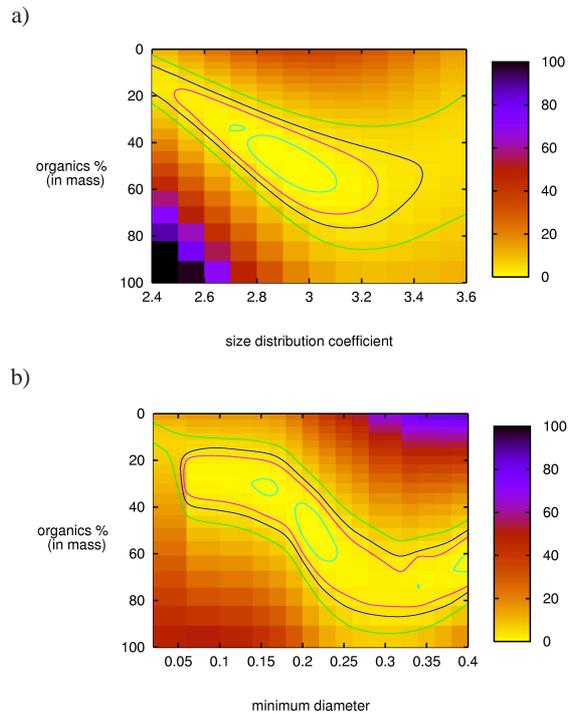

\centering
a) \hspace{0.8\linewidth} \, \\
\includegraphics[width=0.8\linewidth]{chi2p_1.epsf}\\
b) \hspace{0.8\linewidth} \, \\
\includegraphics[width=0.8\linewidth]{chi2p_2.epsf}
\caption{$\chi^2$ maps of the model compared to the polarization data. 
a) $\chi^2$ as a function of the maximum organics mass ratio and 
the small grains size distribution coefficient for a lower diameter of 
\unit{0.22}{\micro\metre} and b) $\chi^2$ as a function of the maximum organics 
mass ratio and the small grains lower diameter for a size distribution power law 
in $\Gamma (a)=a^{-3}$.
The confidence levels curves for 70\%, 90\%, 95\% and 99\% are represented.
}
\label{chi2}
\end{figure}

In Fig.~\ref{chi2} are plotted the root mean square values $\chi^2$ of our model 
compared to the polarization data points, 
calculated following the equation:
\begin{equation}
\chi_{pol}^2=\sum_i \left| \frac{P_{obs}(\alpha_i)-P_{model}(\alpha_i)}{\sigma_P(\alpha_i)} \right|^2
\end{equation}
The plots are shown as 2 dimensional maps 
as a function of the maximum organics percentage in mass of the cloud, and the two other main parameters 
of the model (small grains size distribution coefficient and cutoff diameter). 
The first map shows that the size distribution coefficient for small grains 
(in the 0.2 to \unit{20}{\micro\metre} diameter range) 
plays an important role for the shape of the phase curve, with a minimum located around $-2,95\pm 0,15$,
thus confirming that a realistic size distribution of the particles should have a coefficient close to this value. 
The organics percentage in mass is around 50\%. 
The second plot presents the dependence with the minimum diameter and the organics percentage in mass. 
This shows that the minimum is obtained for a lower cutoff around \unit{0.22}{\micro\metre}. 
The high variation of the function towards higher values of the lower cutoff shows the predominance of the small 
particles in the final shape of the polarization phase curve. However too large lower cutoffs can be ruled 
out because IDPs studies have shown the importance of constituent grains 
in the 0.1 to \unit{0.3}{\micro\metre} size range
as building blocks of interplanetary dust particles.
Upper cutoffs larger than \unit{200}{\micro\metre} do not constrain the $\chi^2_{pol}$
so that the larger particles of the size distribution do not significantly influence the
shape of the polarization phase curve.

\begin{figure}
\centering
\includegraphics[width=0.8\linewidth]{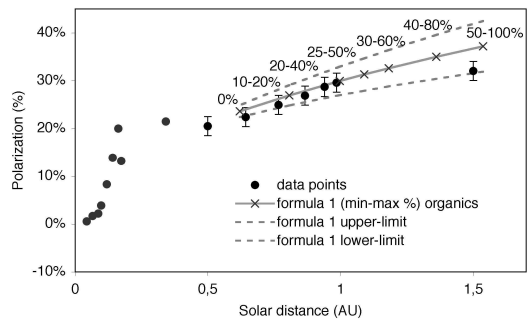}
\caption{Dependence of $P_{\unit{90}{\degree}}$ with $R$. 
Data points ($\bullet$) are compared to the previous IDC model with a silicates-organics  composition
expressed as percent of organics ($\times$) varying with the distance to the Sun.
 Percentage values take into account the possible presence of core-mantle particles. 
Data points are adapted from Levasseur-Regourd et al. \cite{aclr_im01} and 
references therein. The error bars are shown whenever greater than 0.5\%.}
\label{fit_P_R}
\end{figure}

In Fig.~\ref{fit_P_R}, the dependence of $P_{\unit{90}{\degree}}$ with $R$  
deduced from the observations (equation~\ref{var_P})
is compared to the above model of particles
 with an organic material percentage varying with the distance to the Sun
from 50-100\% to 0\% as indicated on the curve. The percentage values correspond to the quantity of
organic material required to reproduce the variation of the observations best fit 
(variation given by the equation~(\ref{var_P})
which is more accurate near 1~AU than near 1.5~AU).
For a given distance to the Sun the percent range corresponds 
to the possible existence of core-mantle particles (Lasue \& Levasseur-Regourd \cite{jl_aclr06}). 
A significant loss of organics is mandatory to explain the decrease of  $P_{\unit{90}{\degree}}$ in the 
1.5 to 0.5~AU range.
The  decrease of $P_{\unit{90}{\degree}}$ with $R$ from 1.5 to 0.5~AU
could thus be related  to a change in the material ratio of the IDC. 
However, other parameters such as the size distribution parameters (cutoff diameters or exponent $s$)
could also change with the distance to the Sun when considering small distances to the Sun 
(in the 0.5 to 0~AU range) where a drastic change in polarization is observed.

\subsection{Discussion}

The light scattering analysis (Fig.~\ref{fit_Palpha}) indicates the presence of two types of  material 
(typically non-absorbing silicates and more absorbing carbonaceous material), 
possibly mixed together in the IDC in the form of 
silicates-core, organics-mantle particles. It confirms previous studies showing that
a significant part of the IDC is generated from cometary dust where these materials 
have been observed in large quantities (see e.g. Greenberg \& Hage \cite{jmg_jih90}). 
It also agrees with infrared observations showing silicates emission features 
(see e.g. Reach et al. \cite{wtr_aa96}, Leinert et al. \cite{cl_pa02}), 
with the analysis of the IDPs collected in the stratosphere of the Earth and micrometeorites collected in ices
(for a review, see Jessberger et al. \cite{ekj_ts01})
and with recent analyses of Stardust samples
(see e.g. Zolensky et al. \cite{mez_tjz06}, Sandford et al. \cite{sas_ja06}).
However it does not rule out the existence of significant contribution from 
dust resulting from asteroidal collisions.

The heliocentric dependence analysis (Fig.~\ref{fit_P_R}) also shows that a change in the composition 
of the IDC, with a decrease of the  absorbing carbonaceous material percentage
when the solar distance decreases, can explain the decrease of $P_{\unit{90}{\degree}}$ from 1.5 to 0.5~AU. 
This interpretation is coherent with previous solar F-corona and zodiacal light data 
showing a probably wide extended zone of degradation of the carbonaceous material 
away from the Sun (Mann et al. \cite{im_ho94}) reaching up to 1.8~AU (Mukai \cite{tm96}). 
It is also in agreement with recent mid-infrared observations suggesting that the silicates emission 
feature  decreases further away from the Sun (Reach et al. \cite{wtr_pm03}).
It finally supports the decrease in the local albedo value of the particles further away from the Sun
 deduced from the observations (Levasseur-Regourd et al. \cite{aclr_im01}).

\begin{figure}
\centering
\includegraphics[width=0.8\linewidth]{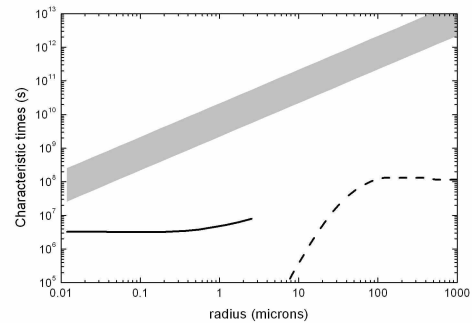}
\caption{Comparison between characteristic migration times and characteristic degradation times of spherical particles 
as a function of their radius. The gray zone corresponds to characteristic migration times of spherical particles 
submitted to the Poynting-Robertson effect with a density between \unit{1000}{\kilogrampercubicmetrenp} (upper limit)
and \unit{100}{\kilogrampercubicmetrenp} (lower limit). The dashed and plain lines correspond to the degradation time of POM
and poly-HCN particles respectively at 1~AU.
}
\label{imageFray}
\end{figure}

%%%%%%%%%%%%%%%%%%%
% rajout H. Cottin et N. Fray
This interpretation is also consistent with observations within cometary atmospheres. It has already been shown 
that solid organic material ejected from cometary nuclei on grains can be degraded when the temperature of the 
grains rises in the coma. The degradation of polyoxymethylene (formaldehyde polymers: (-CH$_{2}$-O-)n, also called POM) 
has been proposed to be the origin of the formaldehyde extended source observed in comet Halley  (Cottin et al. \cite{hc_yb04}).
Such a mechanism is also consistent with the variation of the production of H$_2$CO in comet Hale-Bopp from 0.9~UA to 
4~UA (Fray et al. \cite{nf_yb06}). However POM is a rather fragile compound and its degradation would only contribute to a 
variation of grain composition over a short period of time, inside the coma under thermal effect. From 1.5 to 0.5~AU, 
the characteristic time of POM disappearance is much lower than the one of migration towards the Sun under 
Poynting-Robertson effect. As an example, considering a grain with an effective radius of \unit{100}{\micro\meter} at 1~AU having 
a black body temperature near \unit{280}{\kelvin}, 99\% of POM presumably present in this grain would disappear in about 
\unit{10^{8}}{\second} (covering about $10^{-4}$~UA) whereas its migration time towards the Sun is about \unit{10^{12}}{\second},
 as can be seen in Fig.~\ref{imageFray}.
Thus, POM cannot explain the dependence of $P_{\unit{90}{\degree}}$ with $R$. 

However, other organic compounds, such as HCN polymers (Rettig et al. \cite{twr_tsc92}) have been proposed as being part of 
the solid organic component of cometary grains. Their degradation could be in part responsible for 
the $P_{\unit{90}{\degree}}$ variation. 
Preliminary results on thermal degradation of HCN polymers have shown that this compound is much more resistant than 
POM (Fray et al. \cite{nf_yb04}) and that it starts to decompose only for temperature higher than \unit{450}{\kelvin}. 
Setting this triggering temperature implies that HCN polymer could survive closer to the Sun than POM. 
Other solid carbonaceous compounds should also be considered, such as amorphous carbon or 
carbon nitride, for which the onset decomposition 
temperature is comprised between 600 and \unit{800}{\kelvin} (Zhang et al. \cite{lhz_hg02}). 
The dependence of $P_{\unit{90}{\degree}}$ with $R$ could probably 
be explained by the degradation of a mixture of refractory compounds, for which degradation is triggered at 
different temperatures, depending on the grain size and the heliocentric distance.
%%%%%%%%%%%%%%%%%%%

Although the variation of polarization can be well 
explained in the 1.5 to 0.5~AU range with the composition change (see Fig.~\ref{fit_P_R}), 
it may be noticed that degradation processes may also change the size distribution
but with a limited effect from 1 to 0.1~AU 
(see e.g. Kimura et al. \cite{hk_im98}, Mann et al. \cite{im_hk04}). 
However,  other processes are mandatory to further 
explain the drastic variation of polarization for $R$ below 0.5~AU.
Gail \& Sedlmayr (\cite{hpg_es99}) have shown that compounds such as enstatite and forsterite 
(typically Mg-rich pyroxene and olivine) have their stability limits 
under equilibrium conditions in the temperature region between  \unit{800}{\kelvin} and  \unit{1100}{\kelvin}. 
Such temperatures should be reached for distances $R$ to the Sun  below 0.3 AU 
(see e.g. Kimura et al. \cite{hk_im02}, Mann et al. \cite{im_hk04}). 
The degradation of silicates mineral
in this region could explain the observed variation of the polarization.

The amount of fluffy particles used in our model is of about 20\% in mass. 
This may be related to the fact that unfragmented particles originating from comets 
constitute at least 20\% of the IDC in mass. This is only a lower limit value since 
not all the cometary dust particles 
present a porous structure (H\"orz et al. \cite{fh_rb06}).
Such a result reasonably agrees with previous estimations of the asteroidal over cometary dust ratio
deduced from collisional evolution of asteroids (at least 1/3 of IDC is of asteroidal origin, 
Dermott et al. \cite{sfd_kg96}) 
and from lunar impact analysis (2/3 is of asteroidal origin, Fechtig et al. \cite{hf_cl01}).

%%%%%%%%%%%%%%%%%%%%%%%%%%%%%%%%%%%%%%%%%%%%%%%%%%%
\section{Equilibrium temperature of the particles}

\subsection{Principle of calculations}

The temperature, $T$, of dust particles in thermal equilibrium at a solar distance, $R$,
typically between 1.5 and 0.5~AU, is computed by equalling incident and  emitted energy over 
the ultraviolet, the visible and infrared spectrum, from around 0.1 to \unit{1000}{\micro\metre}
\begin{eqnarray}
\left(\frac{r}{R}\right)^{2}\int_{0}^{\infty} B(\lambda,T_{\rm{S}} ) Q_{\rm{abs}}( a, \lambda ) d \lambda  \nonumber \\
= \varsigma \int_{0}^{\infty}B( \lambda ,T)Q_{\rm{abs}}( a, \lambda ) d \lambda 
\end{eqnarray}
where $r$ is the radius of the Sun, 
$B(\lambda,T)$ is the Planck function, $T_{\rm{S}}$ the solar surface temperature, 
$\varsigma$ the ratio of the emitting surface over $\pi a^{2}/4$ and 
$Q_{\rm{abs}}(a,\lambda)$ the absorption efficiency of a particle with a given optical index 
$m(\lambda)$ (Kolokolova et al. \cite{lk_msh05}). 

Since the absolute value of $T$ at 1~AU is not accurately known,  
 we shall mainly discuss the variation of $T$ with $R$ and tentatively reproduce such 
a variation in the frame of the previous model. 

Calculations are performed for various shapes; 
core-mantle spheres, core-mantle prolate spheroids and typical fractal 
aggregates (BCCA and BPCA) thereof.
The optical indices are taken to be those of astronomical silicates
(Draine \& Lee \cite{btd_hml84}),
Mg-rich pyroxene (as described in 4.1, Dorschner et al. \cite{jd_bb95}),
refractive organics material  (Li \& Greenberg \cite{li_jmg97})
or amorphous carbon (Edoh \cite{oe_83}).

\subsection{Results for the temperature of isolated dust particles}

\begin{figure}
\centering
\includegraphics[width=0.8\linewidth]{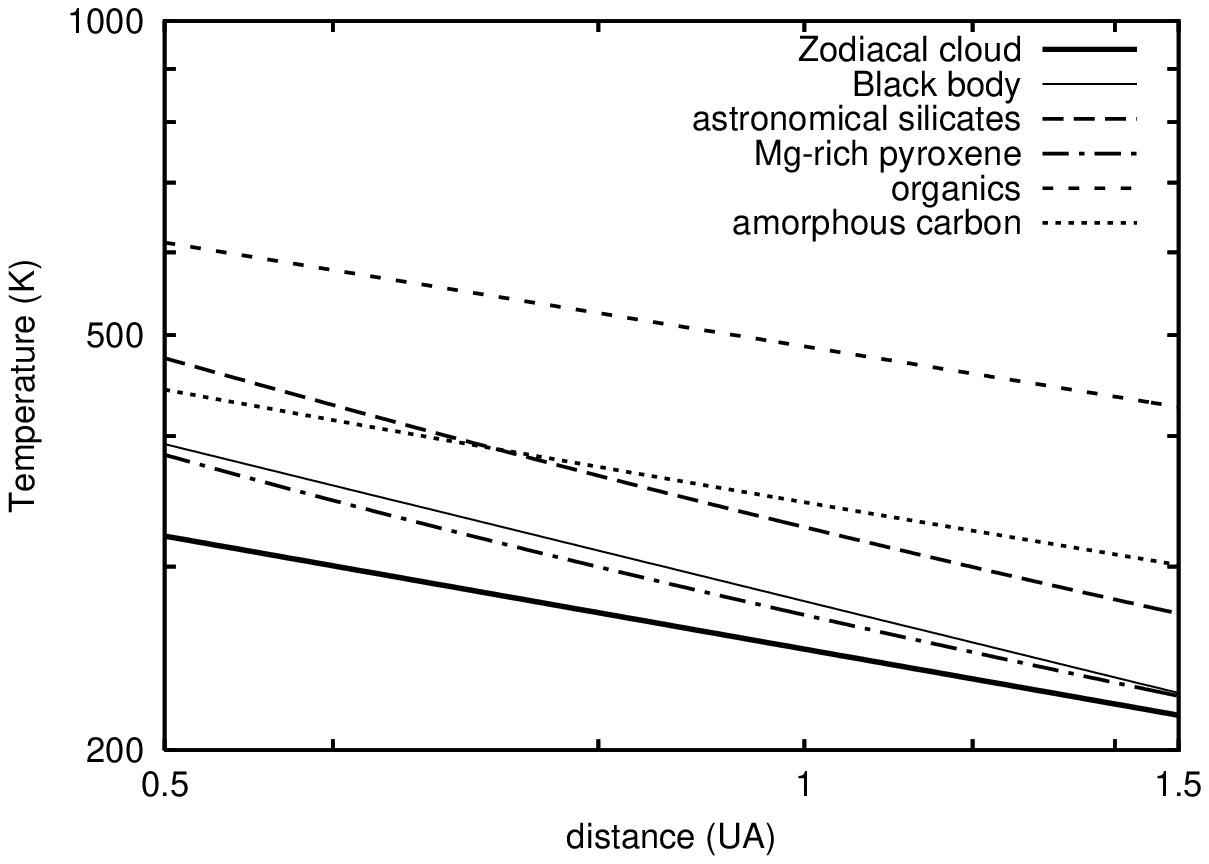}
\includegraphics[width=0.8\linewidth]{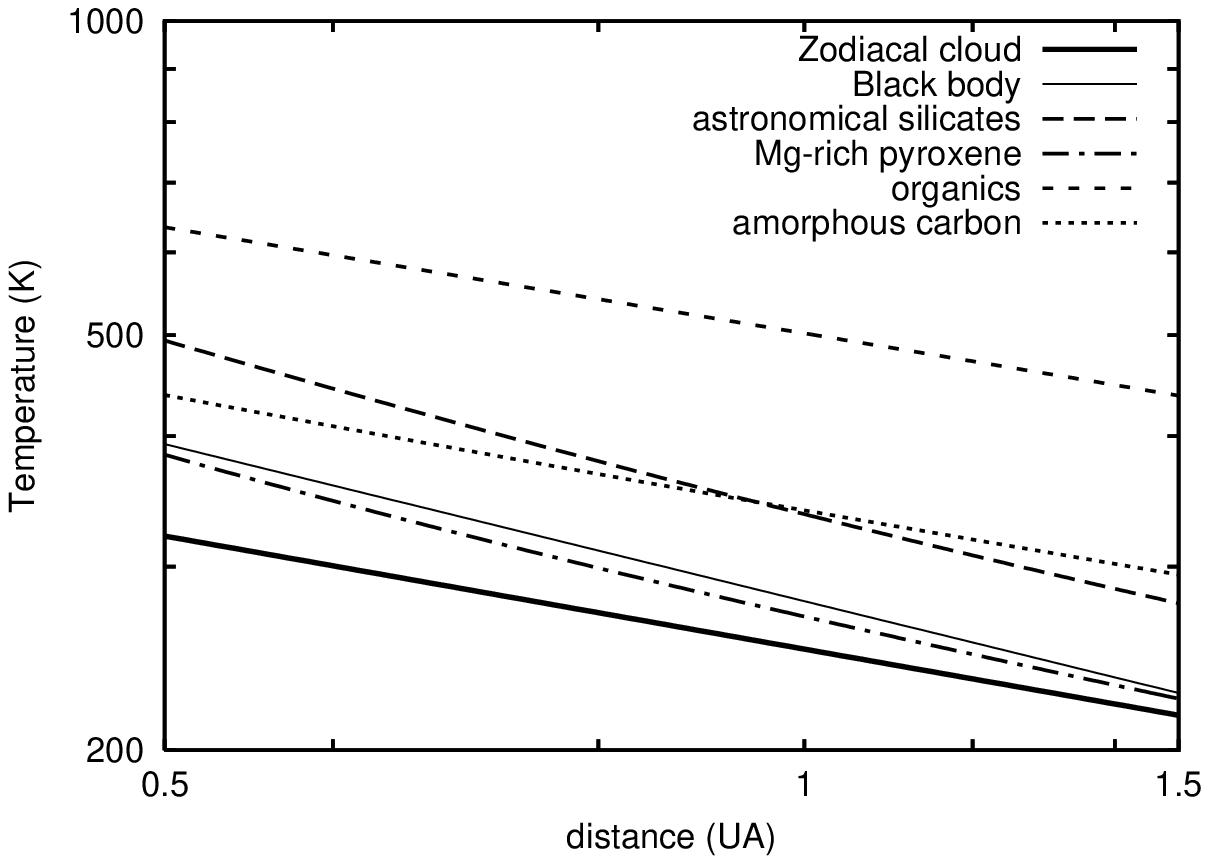}
\includegraphics[width=0.8\linewidth]{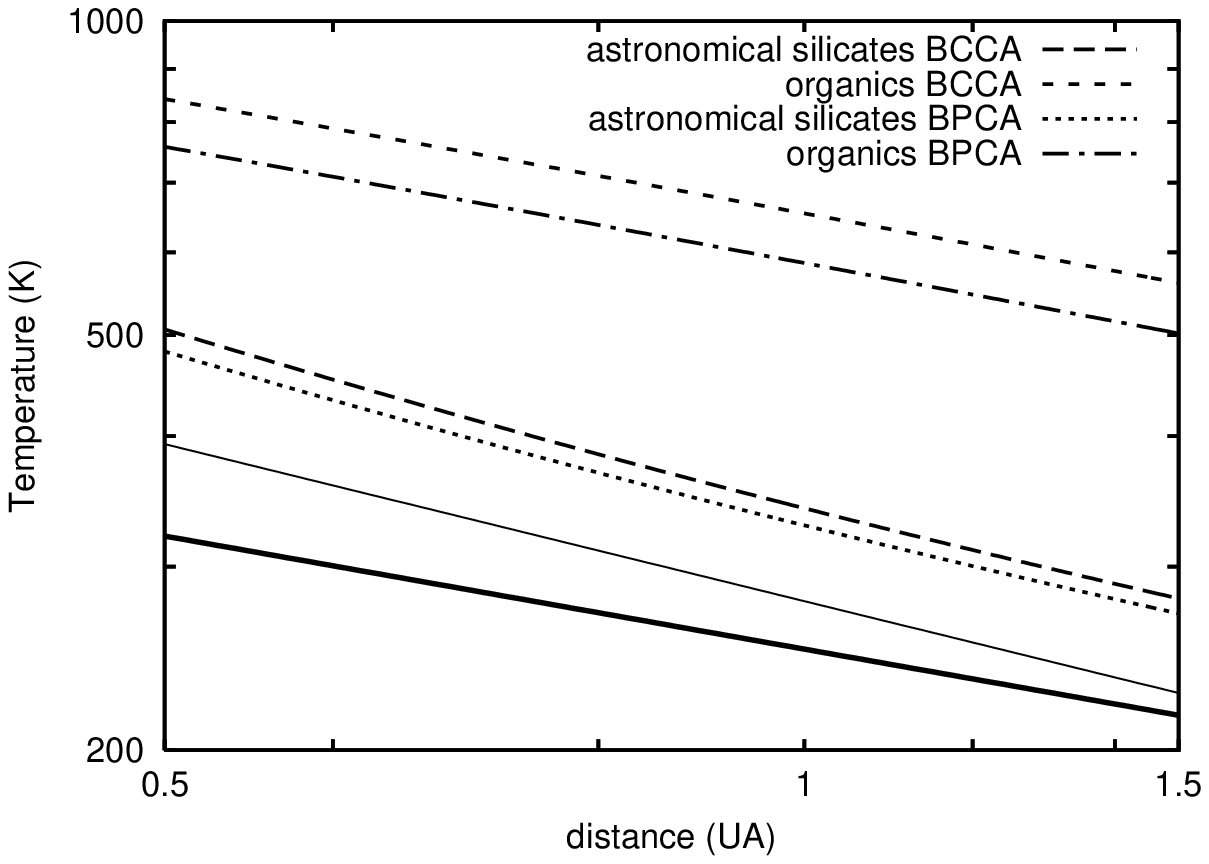}
\caption{Logarithmic plot of the dust temperature in the symmetry surface as a function of the solar distance. 
Temperature inferred from observations (thick solid line) as compared to 
black body temperature (thin solid line) and temperature computed 
for spheres (top), spheroids (middle) 
and BCCA-BPCA aggregates (bottom) of equivalent diameter \unit{1.5}{\micro\metre}. 
A satisfactory trend for the variation of the temperature with the solar distance
is obtained in the case of absorbing organics or carbon material.}
\label{grad_soc}
\end{figure}

Fig.~\ref{grad_soc} presents the variation of the equilibrium temperature with the heliocentric distance
 in logarithmic scale between 1.5 and 0.5~AU for respectively spheres, 
spheroids and 64 spheres BCCA-BPCA aggregates 
(of \unit{1.5}{\micro\metre} equivalent volume sphere diameter meaning constituting spheres of \unit{0.19}{\micro\metre} 
radius) made of astronomical silicates, Mg-rich pyroxene, organics or amorphous carbon material.
The curves are systematically compared to the IDC temperature variation (equation~\ref{var_T}) retrieved from the observations 
(thick solid line) and the black body temperature (thin solid line). 
The absolute value of the temperature is better approximated by black body particles, low-absorbing 
materials (such as Mg-rich pyroxene) and the more compact particles amongst the aggregates. 
This is in agreement with previous works (see e.g. Reach et al. \cite{wtr_pm03} and Kasuga et al. \cite{tk_ty06})
showing that the particles near 1~AU are rather large and compact with a behaviour close to a black body.
However, the temperature-distance factor, $t$, between 0.5 and 1.5~AU  is approximately equal to $-0.33$ for organics, 
$-0.35$ for amorphous carbon which is significantly closer to the observations (equation~(\ref{var_T})) than the values
of $-0.52$ for astronomical silicates and $-0.5$ for Mg-rich pyroxene. 
This suggests that absorbing materials constitute a non negligible component of the IDC 
between 0.5 and 1.5~AU.

The heliocentric power law dependence of the temperature does not change significantly with the shape of the particle 
(spheres, spheroids or BCCA-BPCA aggregates) and is rather defined by the effective diameter and 
the material constituting the particle. The variation of $t$ with
the size of the particles is not accurately known, but should tend towards a black body law when the 
equivalent diameter of the particles increases. 

\begin{figure}
\centering
\includegraphics[width=0.8\linewidth]{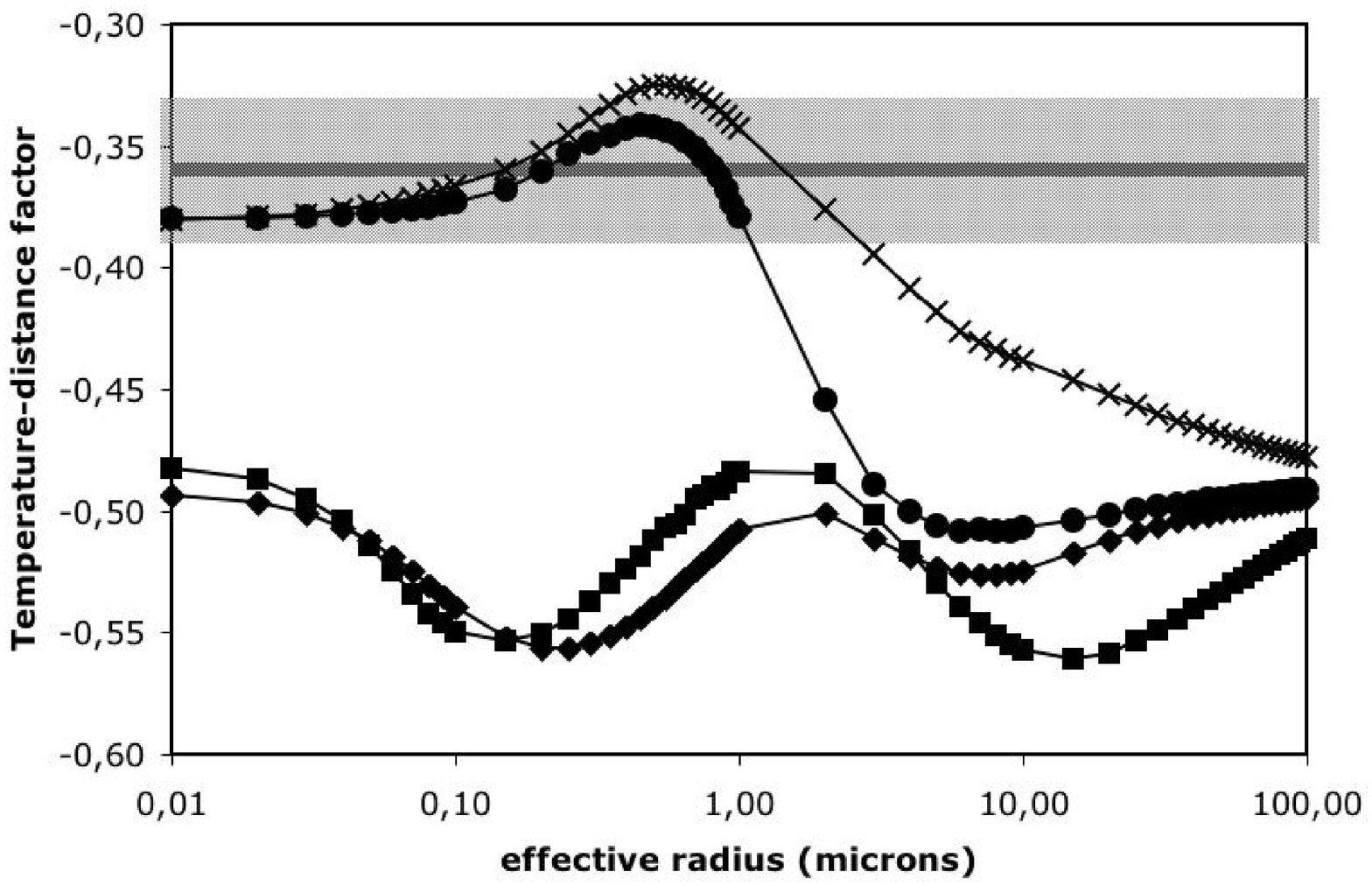}
\includegraphics[width=0.8\linewidth]{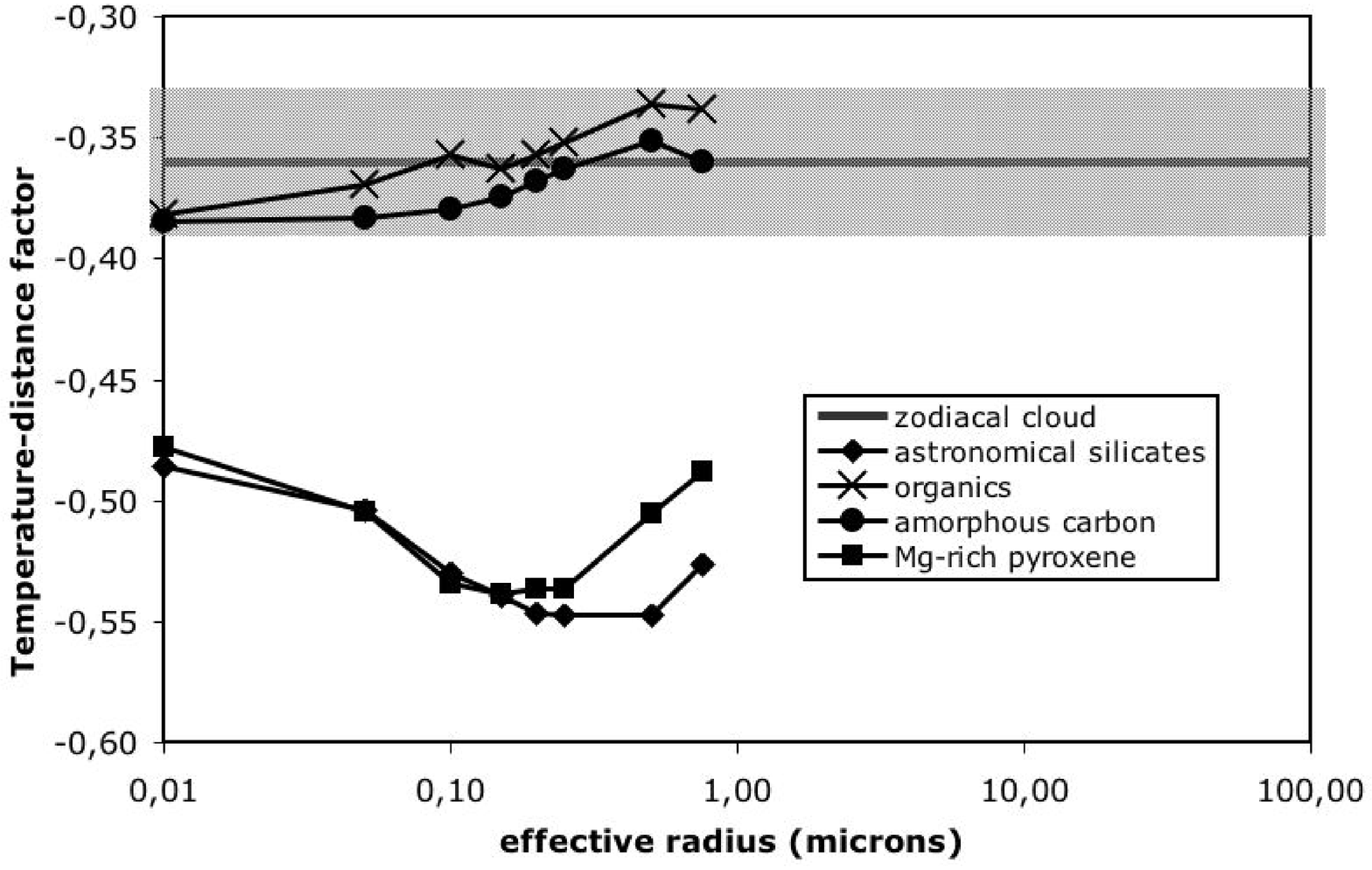}
\caption{Comparison between temperature-distance factors in the symmetry surface and 
between 0.5 and 1.5~AU, as inferred from observations (thick solid line surrounded by gray zone) 
and as calculated for spheres (top) and  
spheroids (bottom) as a function of their equivalent radius.
Calculations are presented for astronomical silicates ($\blacklozenge$), Mg-rich pyroxene 
($\blacksquare$) and absorbing material like  organics ($\times$) and amorphous carbon ($\bullet$) material.} 
\label{grad_size}
\end{figure}

Fig.~\ref{grad_size} shows the comparison between the temperature-distance factors retrieved 
from the observations (the estimated error bar being represented by a gray zone) and 
 computed for spheres and spheroids as a function of the size of the particles.
For an equivalent radius larger than \unit{10}{\micro\metre}, the equilibrium temperature of these particles 
behave like the one of a black body. 
For smaller sizes (\unit{<1}{\micro\metre}), the temperature-distance factor, $t$, of the astronomical silicates and pyroxene 
particles is around the black body value whereas for the (more absorbing) organics and carbon material,
$t$ is equal to the temperature-distance factor retrieved from the observations within the error bars.

\begin{figure}[!t]
\center{\includegraphics[width=0.8\linewidth]{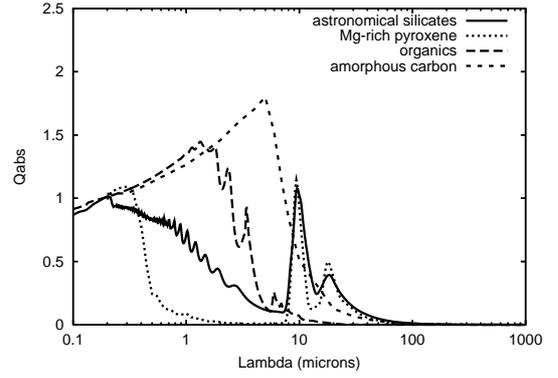}}
\caption{Comparison of absorption efficiency, $Q_{abs}$, for spheres made of different 
materials (astronomical silicates, Mg-rich pyroxene, organics, amorphous carbon) 
as a function of the wavelength for an equivalent 
radius of \unit{0.75}{\micro\metre}.}
\label{qabsM_soc}
\end{figure}

As illustrated in Fig.~\ref{grad_soc}, the temperatures obtained for more absorbing 
materials (organics and amorphous carbon) are higher than those corresponding to less absorbing 
materials (astronomical silicates and Mg-rich pyroxene) for the same equivalent radius. 
This corresponds to the fact that organics or carbon particles absorb more than the silicates in the 
visible and the near infrared part of the spectrum, whereas they emit less in the infrared as can be seen 
in Fig.~\ref{qabsM_soc} that shows the absorption efficiency of spherical particles of radius  
\unit{0.75}{\micro\metre}.

Higher values of the temperature are obtained for fluffier particles: BCCA compared to BPCA aggregates, and 
aggregates compared to equal volume spheroids. This indicates the importance of the size and porosity 
of the particles on their equilibrium temperature. It results from the fact that the more porous 
 an aggregate is, the lower is the contribution of the interaction between constituent particles, and 
the more prominent become the properties of its individual, constituent particles (in agreement with 
Xing \& Hanner \cite{zx_mh97} and Kolokolova et al. \cite{lk_hk07}).

\subsection{Equilibrium temperature of a cloud of spherical particles}

\subsubsection{Principles of the emissivity calculation of the particles}

The emissivity, $\varepsilon_\lambda$, of a cloud of particles with different compositions 
at a given wavelength and distance to the Sun is obtained by summing the Planck function contribution
of each particle (with a temperature, $T$, and an absorption efficiency, 
$Q_{abs}$, depending on its 
size). These considerations are summed up in the following equation from Reach et al. (\cite{wtr_pm03}):
\begin{equation}
\varepsilon_\lambda = \sum_{i} \int \Gamma(a)^{(i)} B_\lambda \left( T^{(i)}(a) \right) \pi a^2 Q_{abs}^{(i)}(a,\lambda)  da
\end{equation}
where the sum over $i$ corresponds to the sum over the different materials constituting the cloud, 
$a$ is the equivalent diameter of the particles and $B_\lambda \left( T^{(i)}(a) \right)$ is the value 
of the Planck function at a given temperature, $T^{(i)}(a)$, and a given wavelength, $\lambda$,
 and $\Gamma(a)$ is the same size distribution of particles than the one deduced from the 
polarization calculations in Part~4.
In order to be able to compare the emissivity with Planck curves, the above integral is 
normalized:
\begin{equation}
< \varepsilon_\lambda > = \sum_{i} \frac{ \int \Gamma(a)^{(i)} B_\lambda \left( T^{(i)}(a) \right) 
 \pi a^2 Q_{abs}^{(i)}(a,\lambda)  da}{\int \Gamma(a)^{(i)} \pi a^2 Q_{abs}^{(i)}(a,\lambda) da}
\end{equation}
$< \varepsilon_\lambda >$ also depends on the distance to the Sun since the temperature of the particles 
varies with this parameter.

\subsubsection{Heliocentric variation of the brightness temperature}

\begin{figure}[!t]
\includegraphics[width=0.8\linewidth]{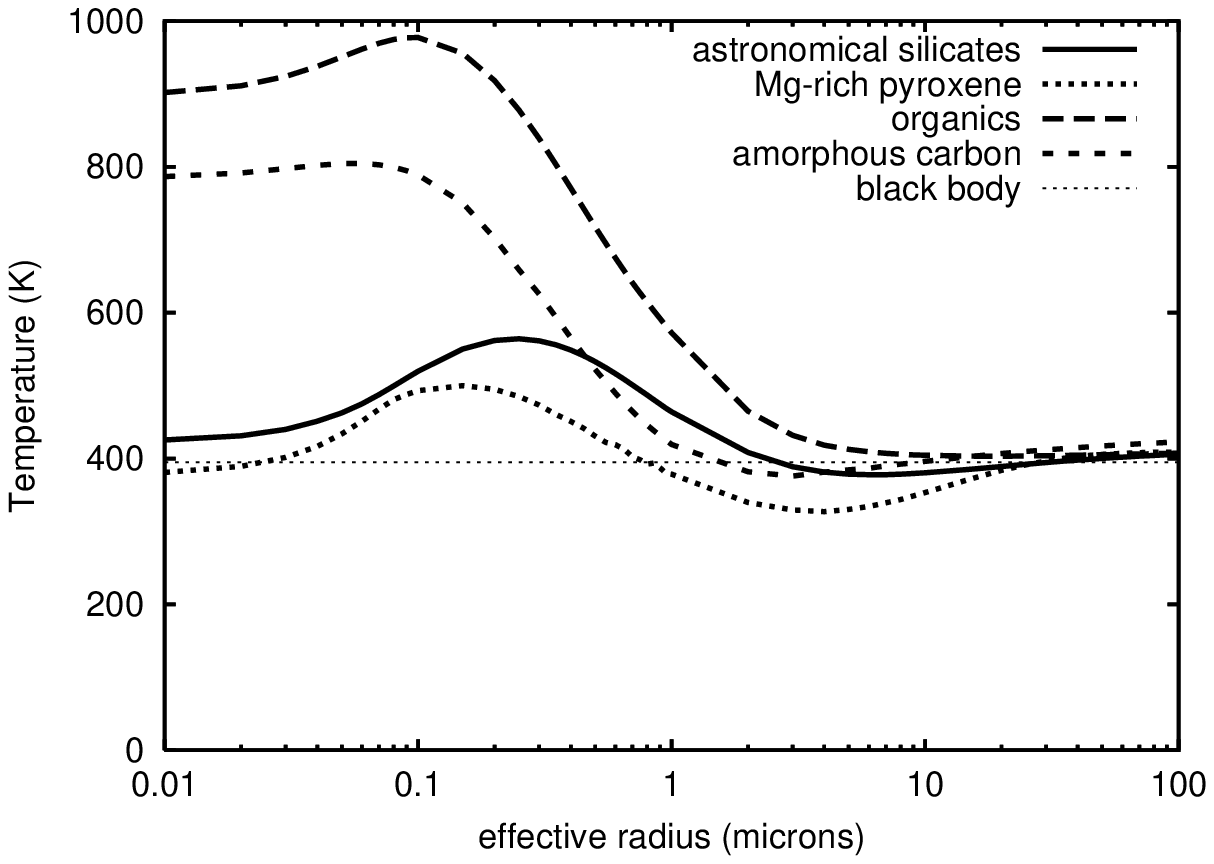}
\includegraphics[width=0.8\linewidth]{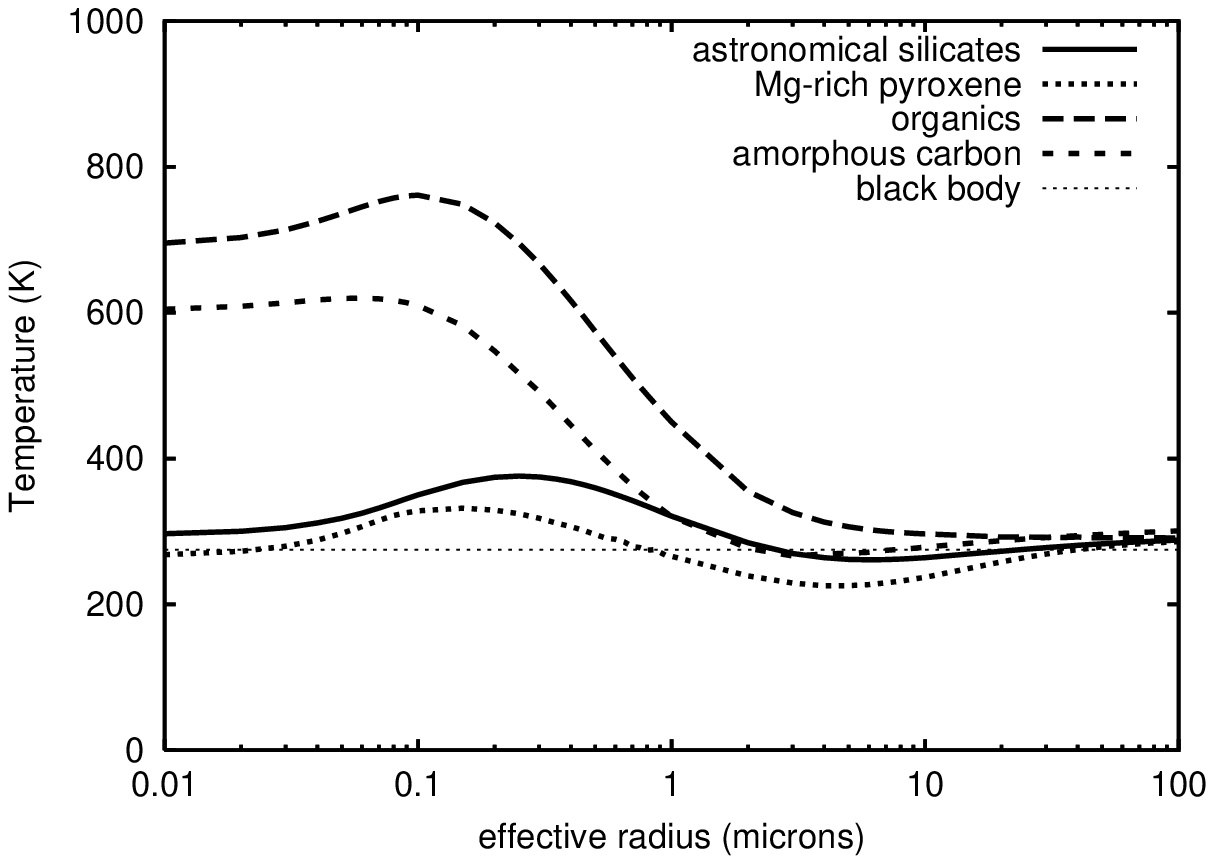}
\includegraphics[width=0.8\linewidth]{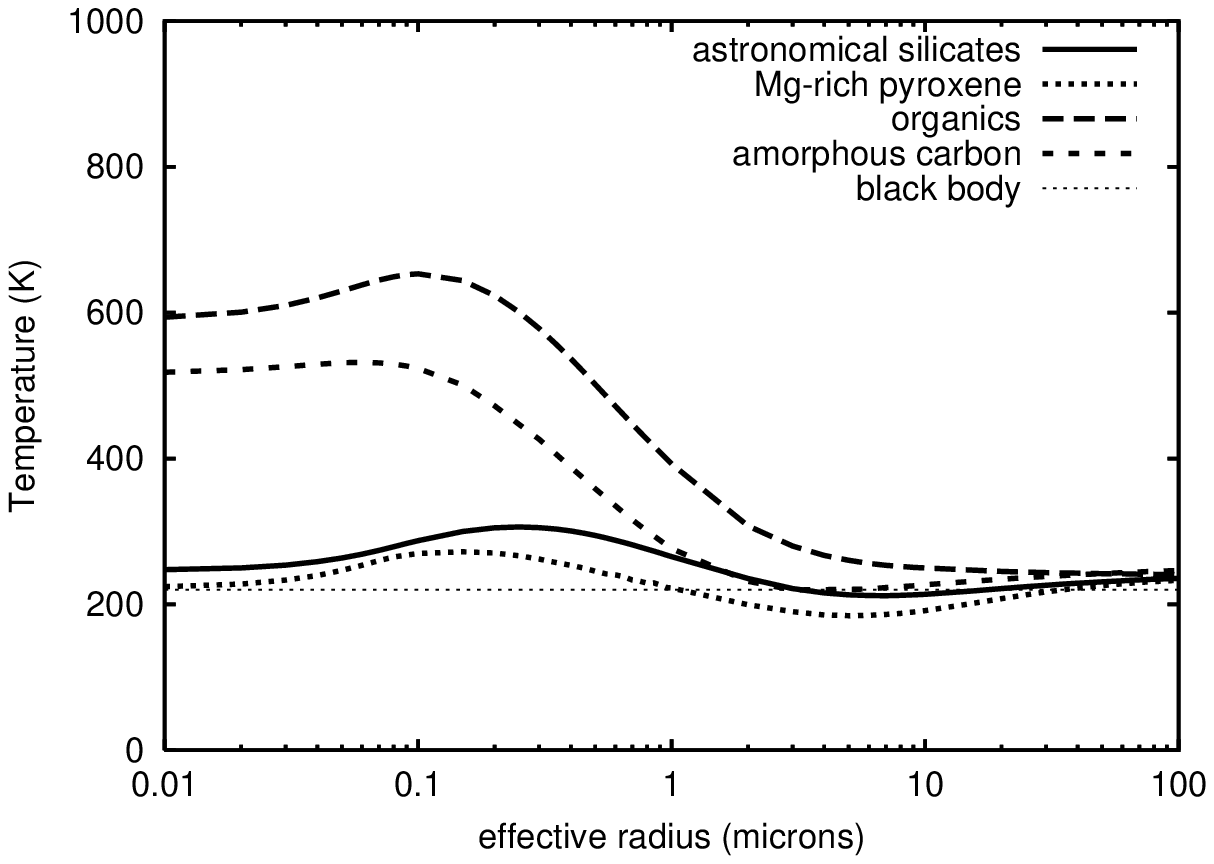}
\caption{Equilibrium temperature of spheres as a function of their radius and their 
constituent materials. The curves are computed for different solar distances:
0.5~AU (top), 1~AU (middle), 1.5~AU (bottom).
}
\label{temp_size}
\end{figure}

The variation of the absorption efficiency with the wavelength 
implies that particles with a radius smaller than \unit{0.1}{\micro\metre} 
absorb and emit very little, and thus do not contribute much to the emissivity function (see e.g. Reach \cite{wtr88}). 
Particles with a size lower \unit{1}{\micro\metre}, 
have generally an equilibrium temperature much higher 
than the black-body one as shown in Fig.~\ref{temp_size}, while particles with a radius larger 
 than \unit{10}{\micro\metre} have an emissivity close to the black body one.

This behavior, already perceptible in the behavior of the temperature-distance factor as a function 
of the particles radius (Fig.~\ref{grad_size}), is confirmed in
Fig.~\ref{temp_size} that shows the variation of the equilibrium temperature of the spheres 
as a function of their radius at a given solar distance (0.5, 1 and 1.5~AU).
Particles with radius smaller than \unit{1}{\micro\metre}, have higher temperatures and 
particles with  radius larger than \unit{10}{\micro\metre} have temperatures similar to a black body.

The approximation of the emissivity curve  $<\varepsilon_\lambda>$ by a Planck curve over the whole 
visible and infrared spectrum (from \unit{\approx 0.1}{\micro\metre} to \unit{\approx 1000}{\micro\metre}) gives 
the equivalent of the brightness temperature of the cloud of particles which corresponds to the 
temperature of the black body that best fits the particles cloud spectrum. 
This temperature calculated at different distances to the Sun gives the temperature-distance factor 
of the particles cloud. Temperatures and temperature-distance factors depending on the distance to the Sun 
are presented in the Tables~\ref{table1},\ref{table2}.

The variation of the brightness temperature with the distance to the Sun depends highly on the material. 
The calculations presented in Table~\ref{table2} confirm that a cloud of silicate particles behave 
like a black body whereas the temperature-distance factor has values lower for a cloud of 
particles constituted of more absorbing materials (organics and amorphous carbon).

The results of the brightness temperature obtained for a mixture of organics and astronomical silicates between 0.5~AU and 
1.5~AU show that a mixture including approximately 50\% of organics  and 50\% of astronomical silicates in mass as 
deduced from the previous light scattering study can explain a behavior significantly different from 
a black body with a temperature-distance factor $t=-0.45$. 
The same temperature-distance factor would also be found if we considered only silicates material at 0.5 AU 
since the brightness temperature of the mixture of silicates and organics is close to the one of silicates 
at this distance (see Table~\ref{table1}).
This result confirms the previous estimations 
of the composition of the IDC from the light scattering observations.

%
%_____________________________________________________________
%                  TABLE TEMPERATURE
%_____________________________________________________________
%

\begin{table*}
\caption{Brightness temperature as a function of the distance to the Sun, $R$, for a cloud of spherical particles
with the size distribution obtained in Part~4.}
\label{table1}      
\centering          
\begin{tabular}{c c c c c c c c }     % 8 columns 
\hline\hline       
                      % To combine 4 columns into a single one 
%HJD & $E$ & Method\#2 & \multicolumn{4}{c}{Method\#3}\\ 
$R$  & observations & astronomical & Mg-rich & organics & amorphous  & 50\% silicates-- & black body\\
  & & silicates & pyroxene & & carbon & --50\% organics &  \\
\hline                    
   0.5 AU & 321$\pm$20 K & 414 K & 371 K & 524 K & 471 K & 408 K & 394 K\\  
   1 AU   & 250$\pm$10 K & 295 K & 255 K & 388 K & 351 K & 298 K & 279 K\\
   1.5 AU & 216$\pm$11 K & 243 K & 210 K & 327 K & 299 K & 249 K & 228 K\\
\hline                  
\end{tabular}
\end{table*}
%
%_____________________________________________________________
%_____________________________________________________________
%             TABLE TEMPERATURE-DISTANCE FACTOR
%_____________________________________________________________
%
\begin{table*}
\caption{Brightness temperature-distance factor, $t$, for a cloud of spherical particles
with the size distribution obtained in Part~4.
$t$ is calculated over three domains (between  0.5 and 1.5 AU).}
\label{table2}      
\centering          
\begin{tabular}{c c c c c c c c }     % 8 columns 
\hline\hline       
                      % To combine 4 columns into a single one 
%HJD & $E$ & Method\#2 & \multicolumn{4}{c}{Method\#3}\\ 
Domain (AU)  &observations & astronomical & Mg-rich & organics & amorphous  & 50\% silicates-- &  black body\\
  & & silicates & pyroxene & & carbon & --50\% organics &  \\
%Domain (AU) &  astronomical silicates & Mg-rich pyroxene & organics & amorphous carbon & 50\% silicates--50\% organics \\
\hline                    
  0.5--1   & $-0.36\pm 0.03$ & $-0.49$ & $-0.54$ & $-0.43$ & $-0.43$ & $-0.45$ &  $-0.5$ \\  
  1--1.5   & $-0.36\pm 0.03$ & $-0.48$ & $-0.49$ & $-0.42$ & $-0.39$ & $-0.44$ &  $-0.5$ \\
  0.5--1.5 & $-0.36\pm 0.03$ & $-0.48$ & $-0.52$ & $-0.43$ & $-0.41$ & $-0.45$ &  $-0.5$ \\
\hline                  
\end{tabular}
\end{table*}
%
%_____________________________________________________________

\subsection{Discussion}

The temperature-distance factor retrieved from the light scattering observations
differs from the black body law. It is possible that the emitting particles may be small and irregular
scatterers, for which the black body approximation cannot be used.
$t$ does not change significantly 
with the shape of the particle (spheres, spheroids or  aggregates)
and is mostly defined by the effective diameter
and the material constituting the particle, in agreement with 
previous temperature calculations for amorphous carbon (Xing \& Hanner \cite{zx_mh97}).
The equilibrium temperature obtained for silicate particles decreases faster with $R$ than for 
absorbing particles, as can be expected from the presence of silicates emission features
in the infrared (Mann et al. \cite{im_ho94}). This can also explain the lower temperature obtained 
for silicates as compared to more absorbing material in Fig.~\ref{grad_soc}.

In agreement with Xing \& Hanner (\cite{zx_mh97}), the temperature obtained for  an aggregate of grains is 
lower than one of its constituent grains but is higher than the temperature of the equivalent volume sphere.
The temperature and polarization thus behave in opposite ways because 
a fluffy aggregate of small grains present values of $t$ close to the one of its
equivalent volume sphere,
whereas it scatters light in a way similar to one of its constituent grains 
as shown from numerical simulations (West \cite{raw91}) and experimental simulations 
(Wurm et al. \cite{gw_hr04}). 

The works of Reach et al. (\cite{wtr88}, \cite{wtr_pm03}) 
use different sphere size distributions 
(power law, interplanetary dust size distribution, Hanner cometary dust size distribution, etc.)
and material similar to the ones we considered here (astronomical silicates, andesite, obsidian, 
amorphous and crystalline olivine and pyroxene, carbon, etc.). 
Their study show a best fit with the observations for a mixture of 10\% carbon and 90\% silicates 
and the interplanetary dust particles size distribution which gives more weight to the particles in the 
10 to \unit{100}{\micro\metre} diameter. 
The conclusions presented in this work fairly agree with their model but has been developed 
specifically to try to interpret the peculiar variation of the temperature with the heliocentric distance.

\section{Conclusions}

The physical properties of the interplanetary dust cloud  
in the near ecliptic symmetry surface are tentatively derived 
from scattered and emitted light observations. Results about the composition of the dust cloud,
the size distribution and the shape of the particles are summarized below.
\begin{enumerate}
\item{Both silicates and (more absorbing) organics materials are necessary to explain the 
local polarization and temperature values retrieved from observations, as well as 
their variation with the solar distance.}
\item{ A good fit of the polarization phase curve available near 1.5~AU 
is  obtained for a realistic particles size distribution
(with a power law $a^{-3}$ for particles with an equivalent diameter, $a$, 
between \unit{0.22}{\micro\metre} and \unit{20}{\micro\metre} and $a^{-4.4}$ for larger particles)
and  a mixture of  silicates and more absorbing  organics materials (between 20\% and 60 \% in mass).}
\item{The upper cutoff of the size distribution is not well constrained with the above mentioned size distribution 
allowing the presence of rather large compact particles.}
\item{The decrease of $P_{\unit{90}{\degree}}$  with the solar distance between 1.5 and 0.5~AU 
is interpreted as a progressive disappearance of the solid carbonaceous compounds 
(such as HCN polymers or amorphous carbon) towards the Sun,
probably linked with the presence of an extended zone of thermal degradation.}
\item{The drastic change of $P_{\unit{90}{\degree}}$ closer to the Sun between 0.5 and 0~AU 
could be explained by other physical processes, as for example
the degradation of silicate materials
or a change in the size distribution possibly favoring  smaller particles 
towards the Sun.}
\item{Unfragmented aggregates of cometary origin contribution to the interplanetary dust cloud 
is of at least 20\% in mass around 1.5~AU.}
\item{The size distribution retrieved from the polarization fit leads to a  temperature variation in $R^{-0,45}$
different from the black body behavior and closer to the observations.}
\item{The variation of the temperature with the solar distance for absorbing materials is closer 
to the observations than the one of non absorbing materials. This behavior is mainly due to particles 
with a diameter smaller than \unit{2}{\micro\metre}.}
\end{enumerate}

\begin{acknowledgements}
This research has been partially funded by CNES. The authors acknowledge fruitful
discussions with J.-B. Renard and helpful comments from an anonymous referee.
\end{acknowledgements}

\end{document}